\documentclass{emulateapj}
\usepackage{apjfonts}
\usepackage{natbib,verbatim,epsfig}

\newcommand{\psra}{PSR~J2140$-$2310A}
\newcommand{\psrb}{PSR~J2140$-$23B}
\newcommand{\msun}{\ifmmode\mbox{M}_{\odot}\else$\mbox{M}_{\odot}$\fi}
\newcommand{\lsun}{\ifmmode\mbox{L}_{\odot}\else$\mbox{L}_{\odot}$\fi}
\newcommand{\rsun}{\ifmmode\mbox{R}_{\odot}\else$\mbox{R}_{\odot}$\fi}
\newcommand{\degrees}{\ifmmode^{\circ}\else$^{\circ}$\fi}
\newcommand{\amin}{\ifmmode^{\prime}\else$^{\prime}$\fi}
\newcommand{\asec}{\ifmmode^{\prime\prime}\else$^{\prime\prime}$\fi}

\slugcomment{Submitted to ApJ}
\shortauthors{Ransom et al.}
\shorttitle{Binary Millisecond Pulsars in M30}

\begin{document}

\title{GBT Discovery of Two Binary Millisecond Pulsars in the Globular
  Cluster M30}

\author{Scott~M.~Ransom\altaffilmark{1,2}, 
  Ingrid~H.~Stairs\altaffilmark{3},
  Donald~C.~Backer\altaffilmark{4},
  Lincoln~J.~Greenhill\altaffilmark{5},
  Cees~G.~Bassa\altaffilmark{6},
  Jason~W.~T.~Hessels\altaffilmark{1},
  Victoria~M.~Kaspi\altaffilmark{1,2,7}}


\altaffiltext{1}{McGill University Physics Dept., Montreal, QC
  H3A~2T8, Canada; ransom@physics.mcgill.ca} 
\altaffiltext{2}{Center for Space Research,
  Massachusetts Institute of Technology, Cambridge, MA 02139}
\altaffiltext{3}{Dept.~of Physics and Astronomy, University of British
  Columbia, 6224 Agricultural Road, Vancouver, BC V6T~1Z1, Canada}
\altaffiltext{4}{Dept.~of Astronomy and Radio Astronomy Laboratory,
  University of California at Berkeley, 601 Campbell Hall 3411,
  Berkeley, CA 94720}
\altaffiltext{5}{Harvard-Smithsonian Center for
  Astrophysics, 60 Garden St., Cambridge, MA 02138}
\altaffiltext{6}{Astronomical Institute, Utrecht University, PO Box
  80\,000, 3508 TA Utrecht, The Netherlands}
\altaffiltext{7}{Canada Research Chair, Steacie Fellow, CIAR Fellow}

\begin{abstract}
  We report the discovery of two binary millisecond pulsars in the
  core-collapsed globular cluster M30 using the Green Bank Telescope
  (GBT) at 20\,cm.  \psra\ (M30A) is an eclipsing 11-ms pulsar in a
  4-hr circular orbit and \psrb\ (M30B) is a 13-ms pulsar in an as yet
  undetermined but most likely highly eccentric ($e>0.5$) and
  relativistic orbit.  Timing observations of M30A with a 20-month
  baseline have provided precise determinations of the pulsar's
  position (within 4\asec\ of the optical centroid of the cluster),
  and spin and orbital parameters, which constrain the mass of the
  companion star to be $m_2\gtrsim0.1\,\msun$.  The position of M30A
  is coincident with a possible thermal X-ray point source found in
  archival {\em Chandra}\ data which is most likely due to emission
  from hot polar caps on the neutron star.  In addition, there is a
  faint ($V_{555}\sim 23.8$) star visible in archival {\em HST}\ F555W
  data that may be the companion to the pulsar.  Eclipses of the
  pulsed radio emission from M30A by the ionized wind from the compact
  companion star show a frequency dependent duration
  ($\propto\nu^{-\alpha}$ with $\alpha \sim 0.4$$-$0.5) and delay the
  pulse arrival times near eclipse ingress and egress by up to
  2$-$3\,ms.  Future observations of M30 may allow both the
  measurement of post-Keplerian orbital parameters from M30B and the
  detection of new pulsars due to the effects of strong diffractive
  scintillation.
\end{abstract}


\section{Introduction}
\label{sec:intro}

Globular clusters (GCs) produce millisecond pulsars (MSPs) at a rate
per unit mass that is up to an order-of-magnitude greater than the
Galaxy\citep[e.g.][]{ka96}.  Due to the relatively large distances of
GCs (several to tens of kiloparsecs), the low intrinsic luminosities
of MSPs, and the fact that most MSPs are members of compact binary
systems, the discovery of new cluster pulsars requires long
observations with the largest radio telescopes and computationally
intensive data analysis.

The discovery of new cluster pulsars is interesting because of the
wide variety of science that can result from using them as sensitive
probes into the natures of the pulsars themselves and the clusters in
which they live.  Recently, cluster pulsars have been used to probe
properties of GCs, such as the mass-to-light ratios in cluster cores
\citep[e.g.][]{fcl+01,dpf+02}, cluster proper motion \citep{fck+03},
and the ionized gas content in 47~Tucanae \citep{fkl+01}.  For binary
pulsars, timing observations have measured relativistic effects such
as the advance of periastron (and therefore the total mass) in the
47~Tuc~H system \citep{fck+03}, and probed the companion winds and
eclipse mechanisms for several known eclipsing MSPs
\citep[e.g.][]{dpm+01a,pdm+03}.  The precise astrometry provided by
MSP timing has allowed the optical identification of several binary
MSP companions \citep[e.g.][]{fpd+01,egc+02}, which is crucial for
determining the nature of the companion stars, and the X-ray
identification of many MSP systems \citep[e.g.][]{gch+02}, which gives
us useful information on pulsar emission and neutron star cooling
mechanisms.  Finally, many theorists have predicted that truly exotic
pulsar systems, such as a pulsar-black hole binary
\citep[e.g.][]{sig03}, will be found in GCs.

After a flurry of GC pulsar discoveries in the 1980s and early 1990s,
the number of known cluster pulsars remained virtually constant
($\sim$35) until 2000 \citep[for a review, see][]{ka96}.  Over the
past several years, however, the art of searching for radio pulsars in
GCs has undergone a renaissance due to the development of very
sensitive (i.e.~low-noise and high-bandwidth) 20-cm receivers
\citep[e.g.][]{swb+96} and the increasing availability of the high
performance computing resources required to conduct sensitive but
specialized searches for binary millisecond pulsars in observations
with durations of several hours \citep*[e.g.][]{jk91, rem02}.  The
Parkes radio telescope has been particularly productive as of late
with the discovery of at least 24 millisecond pulsars in 8 GCs, most
of which are in binaries \citep{clf+00, dlm+01, dpm+01a, pdm+01,
  ran01, dpf+02, lcf+03, pdm+03}.

In the past two years, the recently upgraded Arecibo telescope and the
new 100-m Green Bank Telescope (GBT) have become available, and
several new algorithms have been developed to improve search
sensitivities to binary MSPs in compact orbits \citep*[][]{cha03,
  rce03}.  Using the GBT and one of these new techniques,
\citet{jcb+02} have recently reported the discovery of three new
binary MSPs in M62 \citep[see also][]{cha03}.  With these advances in
mind, we undertook a major survey of rich and/or nearby GCs at 20\,cm
that are visible from Arecibo and/or the GBT.  We are analyzing the
data using modern search algorithms and techniques.  This is the first
in a series of papers describing the results from these observations.
Specifically, we focus on ``First Science'' data from the GBT, taken
in the Fall of 2001 toward 12 GCs.  Additional discoveries ---
including at least one other new MSP in M13 discovered in the GBT data
described below --- and results from the rest of the project will be
presented elsewhere\footnote{An up-to-date catalog of GC pulsars can
  be found at \url{http://www.naic.edu/$\sim$pfreire/GCpsr.html}}.

\section{Search Observations and Data Analysis}
\label{sec:search}

In 2001 September and October, we used the Gregorian focus 20-cm
receiver (1.15$-$1.73\,GHz usable bandwidth with $T_{sys} \sim 25$\,K)
on the GBT to observe 12 different clusters for either 4\,hrs (M2, M4,
M75, M80, M92, and NGC~6342) or 8\,hrs (M3, M13, M15, M30, M79, and
Pal1) each.  Since the GBT beam has a FWHM of nearly 9\amin\ at
20\,cm, single pointings fully covered each cluster.  Samples from two
orthogonal polarizations were transmitted through fiber optic cables
to the observatory control room where they were fed to either one or
two Berkeley-Caltech Pulsar
Machines\footnote{\url{http://www.gb.nrao.edu/$\sim$dbacker}}
\citep[BCPMs;][]{bdz+97}.  The BCPMs are analog/digital filterbanks
which 4-bit sample each of 2$\times$96 channels at flexible sampling
rates and channel bandwidths and which can sum the two polarizations
in hardware if requested.

We observed each cluster using 96$\times$1.4\,MHz channels of 2 summed
polarizations centered at 1375\,MHz and sampled every 50\,$\mu$s.  For
those observations using a second BCPM (which included the
observations of M30), we recorded 96$\times$1.0\,MHz channels with
summed polarizations sampled every 36\,$\mu$s at a center frequency of
1490\,MHz.  In total, these initial search observations generated
$\sim$0.5\,TB of data, which were stored on DLT~IV magnetic tapes.

Reduction of the data took place on a dedicated Beowulf-style cluster
of 52 dual-processor 1.4\,GHz AMD Athlon workstations located at
McGill University.  The available computing power allowed us to
attempt extensive interference removal (see below) and to conduct
fully coherent acceleration searches\footnote{Acceleration searches
  correct either a time series or its Fourier transform to account for
  large apparent period derivatives in the signals from binary pulsars
  in compact orbits \citep[see e.g.][]{jk91}.} for binary pulsars in
all clusters.  In previous GC pulsar projects acceleration searches
were only used on clusters where pulsars --- and therefore the
dispersion measure (DM) to the cluster --- were known {\em a priori},
due to the computational expense of searching over the additional
``acceleration'' parameter.

Since these data were taken as part of the GBT ``First Science''
program, the telescope was still in the early stages of commissioning
and the radio frequency interference (RFI) environment was quite bad.
Each of the observations contained large quantities of both persistent
and transient broadband and narrowband interference.  The very strong
Lynchburg, Virginia airport radar, with a rotational period of
approximately 12\,s, was particularly destructive and effectively
eliminated any sensitivity we should have had to weak pulsars with
periods longer than $\sim100$\,ms.  Due to the challenges involved in
excising this RFI, approximately one quarter of these data remain to
be fully analyzed.

We searched the raw data from each cluster for RFI in both the time
and frequency domains as a function of both BCPM channel and time (in
units of $\sim$10\,s).  Sections of data containing fluctuations
significant at the 4-$\sigma$ level in the frequency domain or
10-$\sigma$ level in the time domain were masked using the running
median of the appropriate BCPM channel.  In addition, we computed and
Fourier transformed non-dispersed topocentric time series in order to
identify strong and obvious terrestrial interference (i.e.~occurring
at ``un-natural'' frequencies such as $11.\bar{1}$\,Hz) which we
ignored in subsequent stages of the analysis.

After applying the interference masks, we de-dispersed the data over a
range of DMs from $\sim$50$-$200\% of the predicted DM of the cluster
\citep[using the galactic electron density model of][]{tc93}, or, for
clusters with known pulsars, from $\sim$90$-$110\% of the average DM
of those pulsars.  The clusters that we observed with the GBT were
chosen partly because their known or predicted DMs were
$\lesssim$100\,pc\,cm$^{-3}$, thereby keeping the dispersive smearing
across each BCPM channel to $\lesssim$0.5\,ms at 20\,cm.  The stepsize
in DM was chosen to produce no more than $\sim$0.2$-$0.3\,ms of
dispersive smearing across the 134.4\,MHz bandwidth.  The final stage
of data preparation involved removing samples from or adding samples
to the time series as appropriate in order to account for the
observatory's motion with respect to the solar system barycenter
\citep[using the DE200 ephemeris of][]{sta82}.

After Fourier transforming each de-dispersed and barycentered time
series and removing the previously identified RFI signals and their
harmonics, we searched the data for pulsations in three steps: 1)
using acceleration searches of the full-length observation to maximize
our sensitivity to isolated or long-period binary pulsars, 2) with
acceleration searches of overlapping $\sim$20\,min and $\sim$60\,min
segments of the observation to improve our sensitivity to pulsars in
compact binaries or which exhibit significant diffractive
scintillation, and 3) using phase-modulation (or sideband) searches
for pulsars in ultra-compact ($P_{orb}\lesssim$2\,hr) binary systems
\citep{rce03}.  For the acceleration searches, we used a modern
Fourier-domain matched-filtering code that includes Fourier
interpolation and the incoherent harmonic summing of the first 1, 2,
4, and 8 harmonics of any potential signal \citep{ran01}.  The
acceleration search code allowed the highest summed harmonic to
linearly drift by up to 170 Fourier frequency bins during the portion
of the observation being searched (lower harmonics drift a smaller
fraction of bins equal to their harmonic number over that of the the
highest harmonic).  We folded the raw data for interesting
acceleration search candidates over a range of nearby periods, period
derivatives, and DMs in order to maximize the signal-to-noise ratio.
For phase-modulation candidates, we used Fourier-domain
matched-filtering of orbital templates as described in \citet{rce03}.

Given the highly variable RFI environment, the often extreme effects
of diffractive scintillation (see \S\ref{sec:flux}), and the
sensitivity losses caused by uncorrected orbital motion for binary
pulsars, it is difficult to place a hard limit on our overall search
sensitivity.  However, based on search simulations, our detections of
known cluster pulsars, and the radiometer equation, we estimate that
we were sensitive to ``normal'' millisecond pulsars with flux
densities in the range $\sim$50$-$100\,$\mu$Jy.  Our sensitivities to
sub-millisecond pulsars and ultra-compact binary MSPs are factors of
$\sim$2$-$3 worse than the above values due primarily to the effects
of dispersive smearing across the relatively wide BCPM channels for
the former and the incoherent nature of the phase-modulation search
technique for the latter.  For slow pulsars ($P\ga0.1$\,s), our
sensitivities were degraded even more due to the devastating effects
of RFI, and in particular, the Lynchburg radar.

\section{Discoveries and Timing Observations}
\label{sec:timing}

One of the most promising clusters in our search was M30 (NGC~7099; $l
= 27\degrees.18$, $b = -46\degrees.84$), due its
``post-core-collapsed'' nature and relatively high central luminosity
density.  M30 has a measured core of $<$2.5\asec\ as determined by
{\em HST}\ \citep{ygs+94}, and an age and distance of 12.3\,Gyr and
$D$=9.0$\pm$0.5\,kpc \citep[apparent distance modulus $m_v$$-$$M_v =
14.90\pm0.05$, $E(\bv)$ = 0.039;][]{cgc+00} respectively as determined
from {\em Hipparcos} data.  Significant evidence has been found of
mass segregation within the cluster \citep{gwy+98}.

\begin{figure}
  \plotone{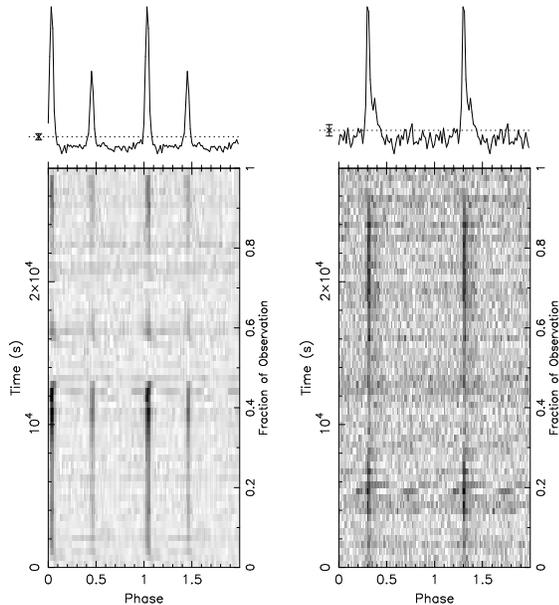}
  \caption{
    Pulse profiles from the 7.8-hr discovery observation of the 11-ms
    binary pulsar \psra\ (M30A, left) and the 13-ms eccentric binary
    pulsar M30B (right).  The Doppler effects of each orbit have been
    removed, and two complete profiles are plotted for clarity.  The
    data were taken with the GBT on 2001 September 9 using the BCPM1
    machine (see \S\ref{sec:search}) at a center frequency of 1370\,MHz
    with an effective time resolution of $\sim$0.12\,ms.  The
    greyscale below the average pulse profiles shows the consistency
    of the pulsed emission as well as some effects due to interference
    as a function of time.  Strong scintillation and portions of at
    least two eclipses of M30A are evident, including dispersive
    delays to the pulse arrival times at eclipse ingress and egress
    (see \S\ref{sec:eclipse}).
    \label{fig:profiles}}
\end{figure}

While performing acceleration searches of $\sim$20\,min segments of
BCPM1 data for M30 taken 2001 September 9, we discovered two highly
variable and highly accelerating pulsar candidates near 11\,ms (M30A;
see \S\ref{sec:m30a}) and 13\,ms (M30B; see \S\ref{sec:m30b}), whose
signal-to-noise ratios peaked at DMs very near 25\,pc\,cm$^{-3}$.
Upon examining the BCPM2 data from the same time period, we confirmed
both of these candidates as pulsars with very high significance.  Soon
thereafter, we determined ``local'' timing solutions for these data
consisting of a compact 4.2-hr orbit for M30A and a high-order
polynomial expansion in spin frequency for M30B which showed each
pulsar visible for $\sim$90\% of the 7.8\,hr observation (see
Fig.~\ref{fig:profiles}).


Our discovery of two MSPs within the 9\amin\ beam of the GBT at 20\,cm
--- at least one of which is within 4\asec\ of the cluster center
($\alpha_{M30} = 21^{\rm h} \;40^{\rm m}\;22\fs16\pm 0\farcs 2$,
$\delta_{M30} = -23\degrees\;10\amin \;47\farcs6\pm 0\farcs 2$; J2000;
see \S\ref{sec:opt}) --- and which have measured DMs
($\sim$25\,pc\,cm$^{-3}$) similar to that predicted for the cluster
using both the \citet[$\sim$23\,pc\,cm$^{-3}$;][]{tc93} and the NE2001
\citep[$\sim$41\,pc\,cm$^{-3}$;][]{cl02} models of the galactic
electron density, firmly establish these as the first known pulsar
members of M30.

In 2002 March, we began a series of monthly 4$-$8\,hr observations of
M30 using the GBT/BCPM1 setup described in \S\ref{sec:search}.  We have
been unable to use the second BCPM machine during these sessions due
to problems with the instrument.  The observations were intended to
establish phase coherent timing solutions for both pulsars and to
allow us to search for new pulsars in M30 whose measured flux
densities were temporarily boosted above our detection limits via
scintillation.  On several occasions, we observed M30 with 48\,MHz of
bandwidth centered at either 575\,MHz or 820\,MHz in order to
determine the DM of the pulsars more accurately, take advantage of a
better RFI environment (particularly for 820\,MHz), and to search for
pulsars with spectral indices too steep to be detected near 1400\,MHz.

\section{Pulsar J2140$-$2310A (M30A)}
\label{sec:m30a}

For pulsar M30A we folded the raw data from the timing observations
modulo the predicted pulse period given our best orbital ephemeris and
accumulated the resulting pulse profile until enough signal was
available to allow an accurate measurement of the phase of the
pulsation.  We measured the pulse phase by cross-correlating \citep[in
the Fourier domain;][]{tay92} the observed profiles with a high
signal-to-noise template profile determined during a period of
scintillation-heightened flux density.  Absolute times-of-arrival
(TOAs) resulted by referencing the phase measurement to the start time
of each observation as recorded from the observatory clock.  We later
used {\sc TEMPO}\footnote{\url{http://pulsar.princeton.edu/tempo}} to
correct these times to the UTC(NIST) time standard with data from the
Global Positioning System and to transform them to the solar-system
barycenter using the DE200 planetary ephemeris \citep{sta82}.

\begin{figure}
  \plotone{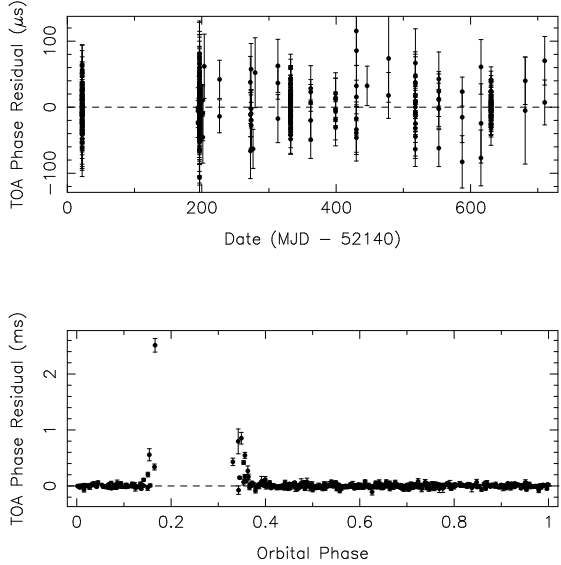}
  \caption{
    (Top) Pulse phase residuals for pulsar M30A as determined after
    fitting for the timing model shown in Table~\ref{tab:M30A} using
    {\tt TEMPO}.  All measured TOAs occurring between orbital phases
    0.12$-$0.38 were excluded from the fit in order to minimize
    systematic effects caused by the pulse delays during eclipse
    ingress/egress.  The RMS residual for the 408 TOAs is
    23.7\,$\mu$s.  (Bottom) All measured phase residuals plotted as a
    function of orbital phase.  For a circular orbit, eclipses are
    expected to occur during superior conjunction at a phase of 0.25.
    Eclipse delays of up to several ms are evident during eclipse
    ingress and egress (see \S\ref{sec:eclipse}).
    \label{fig:resid}}
\end{figure}

Using {\sc TEMPO}, we then fit the TOAs iteratively to a model
incorporating the pulsar position, DM, spin period ($P$), period
derivative ($\dot P$), and the Keplerian orbital parameters for a
circular orbit; the projected semi-major axis ($x\equiv a\sin i/c$),
the orbital period ($P_{orb})$, and the time of the ascending node
($T_{asc}$).  Attempts to fit for eccentricity resulted in a 95\%
confidence upper limit of $e<1.2\times 10^{-4}$.  The timing model
produced RMS residuals of $\sim$24\,$\mu$s (see Fig.~\ref{fig:resid})
and resulted in the parameters shown in Table~\ref{tab:M30A}.

\begin{deluxetable}{lr}
\footnotesize
\tablecaption{Parameters for \psra \label{tab:M30A}}
\tablewidth{0pt}
\tablehead{\colhead{Parameter} & \colhead{Value}}
\startdata
Right Ascension, $\alpha$ (J2000)           & $21^{\rm h}\;40^{\rm m}\;22\fs40610(46)$ \\
Declination,     $\delta$ (J2000)           & $-23\degrees\;10\amin\;48\farcs7936(97)$ \\
Dispersion Measure (pc\,cm$^{-3}$)           & 25.0640(41) \\
Pulsar Period, $P$ (ms)                     & 11.0193290688805(67) \\
Period Derivative, $\dot P$ (s/s)           & -5.181(20)$\times$10$^{-20}$ \\
Epoch (MJD)                                 & 52162.0 \\
Orbital Period, $P_{orb}$ (days)            & 0.17398746418(34) \\
Projected Semi-Major Axis, $x$ (lt-s) & 0.2349416(48) \\
Eccentricity, $e$                           & $<$1.2$\times$10$^{-4}$ \\
Epoch of Ascending Node, $T_{\rm ASC}$ (MJD)& 52161.94552243(68) \\
Span of Timing Data (MJD)                   & 52162$-$52850 \\
Number of TOAs                              & 408 \\
Weighted RMS Timing Residual ($\mu$s)       & 23.7 \\
Flux Density at 1400\,MHz, $S_{1400}$ (mJy) & 0.08(3) \\
\cutinhead{Derived Parameters}
Mass Function, $f_1$ (\msun)                & 0.000459967(28) \\
Minimum Companion Mass, $m_2$ (\msun)       & $\geq$\,0.10 \\
Radio Luminosity, $L_{1400}$ (mJy\,kpc$^2$) & 6.5(2.5) \enddata
\tablecomments{Numbers in parentheses represent the formal 2$\sigma$
  uncertainties in the last digit as determined by {\tt TEMPO}, after
  scaling the TOA uncertainties such that $\chi_{\nu}^2/\nu = 1$.  The
  values for $f_1$ and $m_2$ were derived assuming a pulsar mass of
  1.4\,\msun.}
\end{deluxetable}

\subsection{Cluster Accelerations \label{sec:accel}}

It has been well established \citep[e.g.][]{phi92b} that at least four
different effects can contribute at the few percent or greater level
to the measured spin-down rate, $\dot P$, of a MSP in a GC.  Typically
parameterized in terms of ``accelerations,'' and ignoring the possible
effects of a nearby star or planet perturbing the system, the measured
acceleration of a pulsar is
\begin{equation}
  \label{eqn:accels}
  \frac{\dot P}{P} = \frac{\dot{P_o}}{P} +
  \frac{a_{\ell, \rm GC}}{c} + \frac{a_{\ell, \rm Gal}}{c} + 
  \frac{a_{\rm PM}}{c},
\end{equation}
where $\dot{P_o}$ is the intrinsic spin-down of the pulsar, $a_{\ell,
  \rm GC}/c$ and $a_{\ell, \rm Gal}/c$ are the line-of-sight
accelerations caused by the gravitational potentials of the cluster
and the Galaxy, and $a_{\rm PM}/c = \mu^2D/c$ is the apparent
acceleration caused by the transverse Doppler effect where $\mu$ is
the pulsar's measured proper motion and $D$ is its distance
\citep{shk70}.  Typically, the intrinsic spin-down and cluster
acceleration terms are of the same order, while the Galactic and
proper motion terms contribute at the 10\% level or less.

In the case of M30 and M30A, we know that $\dot P/P=-4.7\times
10^{-18}$\,s$^{-1}$ and can estimate $a_{\ell, \rm Gal}/c \simeq
-8.5\times10^{-19}$\,s$^{-1}$ assuming a spherically symmetric Galaxy
with a flat rotation curve \citep{phi93} and $a_{\rm PM}/c \simeq
1.3\times10^{-18}$\,s$^{-1}$ given the 7.8\,mas\,yr$^{-1}$ of proper
motion measured for M30 by \citet*{dgv99}.  To an accuracy of
$\sim$10\%, \citet{phi92b, phi93} showed that
\begin{equation}
  \label{eqn:maxgc1}
  \max\frac{\left|a_{\ell,\rm GC}\right|}{c} \simeq 
  \frac{1.1G\overline{\Sigma}(<\Theta_{\perp})}{c},
\end{equation}
where $\overline{\Sigma}(<\Theta_{\perp})$ is the projected surface
mass density within the radial position of the pulsar in the cluster,
$\Theta_\perp$ ($\sim 3.6$\arcsec\ for M30A).  If the projected
surface luminosity density is available from optical measurements,
then we can use Equation~\ref{eqn:maxgc1} to place a lower limit to
the projected mass-to-light ratio of the cluster within the radius of
the pulsar position.

Using the power-law relation for the total-light $V$-band surface flux
density in the core of M30 from \citet{gwy+98}, we estimate the
surface luminosity density within the radius of the position of M30A
to be $\sim2.0\times10^4$\,\lsun\,pc$^{-2}$.  Substituting and
rearranging Equation~\ref{eqn:accels} implies that $M/L\gtrsim
0.51\,\msun/\lsun$ in the core of M30, which is typical for GCs.

\citet{phi92b, phi93} also showed that to within 10\% if $\Theta_\perp
< 2\Theta_c$,
\begin{equation}
  \label{eqn:maxgc2}
  \max\frac{\left|a_{\ell,\rm GC}\right|}{c} \simeq
  \frac{3v_\ell^2(0)}{2cD\sqrt{\Theta_c^2+\Theta_\perp^2}},
\end{equation}
where $\Theta_c$ is the core radius of the cluster \citep[which we
have assumed to be $\sim$1.8\arcsec\ based on the range quoted
by][]{ygs+94}, and $v_\ell(0)$ is the one-dimensional velocity
dispersion for the cluster core \citep[9.4$\pm$2.5\,km\,s$^{-1}$ for
M30 from][]{gpw+95}.  Since the measured $\dot P/P$ is negative, the
intrinsic ``acceleration'' must be
\begin{equation}
  \frac{\dot{P_o}}{P} <
  \max\frac{\left|a_{\ell,\rm GC}\right|}{c} + \frac{\dot{P}}{P} -
  \frac{a_{\ell,\rm Gal}}{c} - \frac{a_{\rm PM}}{c}.
\end{equation}
If we account for the 1-$\sigma$ error bars in $D$, $v_\ell(0)$, and
the 10\% accuracy of Equation~\ref{eqn:maxgc2}, for M30A, $\dot P_o /P
< 1.4\times10^{-16}$\,s$^{-1}$.  Similarly, the surface magnetic field
strength of M30A is $B_s = 3.2\times10^{19}(P\dot P_o)^{1/2}$\,G $<$
$4.2\times10^{9}$\,G, the characteristic age $\tau_c = P/(2\dot P_o) >
1.1\times10^{8}$\,yr, and the spin-down luminosity $\dot E =
4\pi^2I\dot P_o/P^3 < 4.6\times10^{34}$\,erg\,s$^{-1}$ assuming the
canonical $I=10^{45}$\,g\,cm$^2$.  Each of these values is typical for
MSPs.

\subsection{Pulsed Flux \label{sec:flux}}

\begin{figure}
  \plotone{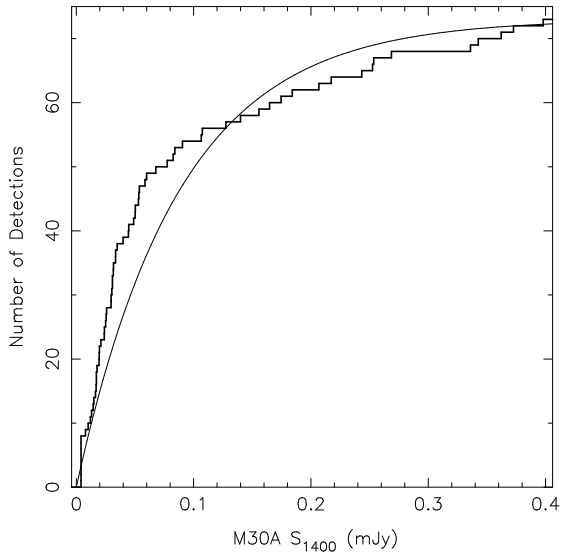}
  \caption{
    A cumulative plot of 73 measurements of the 20-cm flux density of
    pulsar M30A.  Each measurement comprised an hour of integration
    time with the BCPM/GBT setup as described in \S\ref{sec:search}.  The
    thick curve shows a cumulative exponential distribution with an
    average of 87\,$\mu$Jy.  The minimum detectable flux density in
    1\,hr was $\sim$16\,$\mu$Jy.  During six of these 1-hr
    integrations that occurred outside of predicted eclipse times, the
    pulsar was undetected.  For these points, we assumed a flux
    density of half the minimum detectable value, or 8\,$\mu$Jy.
    \label{fig:scint}}
\end{figure}

The monitoring observations have revealed significant diffractive
scintillation \citep[e.g.][]{ric77} which results in large-scale
variations of the measured flux density from M30A over time scales of
one to several hours, and over bandwidths of 50$-$100\,MHz.  By
integrating the pulsed signal above the average off-pulse levels and
comparing the measured noise variance with that described by the
radiometer equation, we have estimated the 20-cm flux density
($S_{1400}$) to an accuracy of $\sim$30\% during 73 1-hour
sub-integrations which occurred during un-eclipsed portions of the
pulsar's orbit.  We did not detect the pulsar during four of the
sub-integrations, so for those epochs we assumed a flux density of 1/2
that of the weakest definite detection we have for the pulsar
($\sim$16\,$\mu$Jy), similar to the procedure used by \citet{clf+00}.
The measurements, displayed as a cumulative plot in
Fig.~\ref{fig:scint}, show the expected exponential distribution for
strong diffractive scintillation \citep[e.g.][]{ric77} with an average
flux density of $\sim$0.08(3)\,mJy.  We made similar measurements for
8 1-hour intervals at 575\,MHz and 10 1-hour intervals at 820\,MHz and
found flux densities of approximately 0.13(4)\,mJy and 0.12(4)\,mJy
respectively.  These measurements correspond to a rather flat radio
spectral index for this pulsar of $-$0.6$_{-0.7}^{+0.5}$.  With such a
low average flux density, it is not surprising that M30A (and the
less-luminous M30B; see \S\ref{sec:m30b}) was not detected in earlier
imaging surveys \citep*[using the VLA]{hhb85} or pulsation searches
\citep[using the 76-m Lovell Telescope at Jodrell Bank]{bl96} of the
cluster.

At the distance of M30, the 20-cm flux density for M30A corresponds to
a radio luminosity $L_{1400}$ = 6.5(2.5)\,mJy\,kpc$^2$.  Comparing
this number to 40 MSPs in the most recent ATNF pulsar catalog
(Manchester et al., in
prep)\footnote{\url{http://www.atnf.csiro.au/research/pulsar/psrcat/}}
with measurements of $L_{1400}$, 11 catalog MSPs with measurements of
$L_{400}$ and assuming a typical spectral index for each of $-$1.6
\citep{lylg95}, and the 14 MSPs in 47~Tucanae with estimates of
$L_{1400}$ in \citet{clf+00}, we find that M30A is in the most
luminous $\sim$30\% of known MSPs, and more than twice as luminous as
the median $L_{1400}$ of $\sim$2.7\,mJy\,kpc$^2$.

The lower frequency observations of M30A have revealed that the
amplitude of the interpulse (as measured with respect to the main
pulse) increases with decreasing radio frequency.  While the
separation in phase at each of the three observing bands
1400/820/575\,MHz seems to be constant at $\sim$42\%, the ratio of the
flux contained in the main pulse to that in the interpulse is
approximately 1.9/1.2/1.0 at 1400/820/575\,MHz, respectively.  Such
frequency dependent evolution of pulse component amplitudes --- while
the pulse phases of the components remain constant --- seems to be a
common occurrence among pulsars \citep[see e.g.][and references
therein]{kll+99}.

\subsection{Eclipse Properties\label{sec:eclipse}}

M30A exhibits consistent and total eclipses at 1400, 820, and 575\,MHz
for approximately 20\% of its orbital period centered on superior
conjunction (i.e.~orbital phase 0.25, see Figs.~\ref{fig:profiles} and
\ref{fig:resid}).  The presence of eclipses implies that the orbit is
significantly inclined ($i>30\degrees$) and therefore the mass of the
companion star is almost certainly in the range $0.10 \la m_2 \la
0.21$\,\msun.  As such, M30A ($P=11.0$\,ms, $P_{orb}=0.174$\,d) is
very similar to the three other eclipsing MSPs with 0.1$-$0.25\,\msun\ 
companions and orbital periods of a few hours, all of which are
located in GCs: PSR~B1744$-$24A in Terzan~5 \citep[$P=11.6$\,ms,
$P_{orb}=0.076$\,d;][]{ljm+90b}, PSR~J0024$-$7204W in 47~Tucanae
\citep[$P=2.35$\,ms, $P_{orb}=0.133$\,d;][]{clf+00}, and
PSR~J1701$-$3006B in M62 \citep[$P=3.59$\,ms,
$P_{orb}=0.145$\,d;][]{pdm+03}.


Scintillation and pulse delays at eclipse ingress and egress (see
below) make the determination of the exact duration of eclipses as a
function of observing frequency difficult.  From a single good
820\,MHz observation, we measure the eclipse duration to be
0.23(2)\,$P_{orb}$, while for four good observations near 1400\,MHz,
the eclipse duration is 0.18(1)\,$P_{orb}$.  This corresponds to a
frequency dependence of the eclipse duration $\nu^{-\alpha}$ with
$\alpha \sim 0.4$$-$0.5, very similar to that measured by
\citet{fbb+90} for PSR~B1957$+$20.  Such a value is also consistent
with the cyclotron absorption based eclipse model of \citet*{kmg00},
where the absorption occurs in the magnetosphere of the degenerate
companion star which gets continuously injected with relativistic
particles from the pulsar wind.

During times of increased signal-to-noise ratio due to scintillation,
we observe pulse phase delays of up to 2$-$3\,ms during eclipse
ingress and egress which imply the presence of an additional electron
column density of $N_e\sim7.4\times10^{17}\Delta_{t,\rm ms}\nu_{\rm
  GHz}^2$\,cm\,$^{-2}$ in the eclipse region, where $\Delta_{t,\rm
  ms}$ is the pulse delay in ms, and $\nu_{\rm GHz}$ is the observing
frequency in GHz (see Fig.~\ref{fig:resid}).  For our observations of
M30A, this corresponds to $N_e\ga 2\times 10^{18}$\,cm\,$^{-2}$.
Since the orbital separation of the system is $a\sim1.4$\,\rsun\, (for
$i \ga 60$\degrees) an eclipse duration of $\sim$20\% of the orbital
period corresponds to a physical size of the eclipsing region $R_E$ of
$\sim$0.9\,\rsun.  The additional electron density near the eclipse
boundaries is therefore $\ga3\times 10^{7}$\,cm\,$^{-3}$ which
corresponds to a plasma density of $\rho_E\ga5\times
10^{-17}$\,g\,cm\,$^{-3}$ if fully ionized.

As in the cases of the other eclipsing MSPs with low-mass companions,
the physical sizes of plausible companion types is much smaller than
$R_E$.  Assuming $i=60\degrees$ and therefore
$m_{2,60\degrees}=0.12$\,\msun, hydrogen and helium white dwarfs (WDs)
have radii of $\sim$0.08\,\rsun\ or $\sim$0.03\,\rsun\ respectively
\citep{st83}, while a zero-age main sequence star would have a radius
of $\sim$0.13\,\rsun.  Even the Roche lobe radius $R_L \sim
0.26$\,\rsun\ \citep[e.g.][]{egg83} is several times smaller than
$R_E$, implying that the eclipses must be due to wind- or
magnetosphere-related activity at several times the companion radius
\citep{fbb+90,sbl+96,kmg00}.  If this wind is emitted isotropically,
the plasma density at the eclipse boundaries corresponds to a mass
loss rate $\dot M = 4\pi R_E^2\rho_Ev_w$, where $v_w$ is the velocity
of the ionized wind from the companion.  If we assume
$v_w\sim10^{8}$\,cm\,s$^{-1}$, which is the order of the escape
velocity from the surface of a presumed WD companion, the
corresponding mass loss rate is $\dot M \sim 4\times
10^{-12}$\,\msun\,yr$^{-1}$.  As has been found with the other
eclipsing systems, unless the ionized fraction of the eclipsing wind
is small ($\la0.4$), the companion star will not be ablated in a
Hubble time.

The nature of the eclipse mechanism for M30A is currently difficult to
constrain given that we know neither the true spin-down luminosity of
the pulsar nor the nature of the companion star.  Additionally, the
low average flux density of M30A has allowed only a few good
measurements of the eclipse ingress and egress.  However, it does seem
unlikely that the eclipses are caused by dispersive smearing of the
pulses, since the excess $N_e$ that we measure near the eclipse
boundaries would cause $\la$1\,ms of smearing across the bandwidth of
our observations and would therefore not smear the pulse enough for it
to appear eclipsed.  While our initial measurement of the frequency
dependence of the eclipse duration seems to be in reasonable agreement
with the cyclotron absorption model of \citet{kmg00}, better
constraints on the eclipse mechanism will have to await higher
sensitivity observations as well as good measurements of the eclipse
properties at several other observing frequencies.

\subsection{Optical Observations\label{sec:opt}} 

We have attempted to identify the companion of M30A in archival
\emph{HST}/WFPC2 observations of M30. We have used the observations in
the F336W, F439W and F555W filters (hereafter $U_{336}$, $B_{439}$ and
$V_{555}$) of GO-5324 (1994 March 1) and the observations in the
$U_{336}$, $V_{555}$ and F814W (hereafter $I_{814}$) filters of
GO-7379 (1999 May 31 and June 1). The exposure times of these
observations are 200\,s in $U_{336}$, 80\,s in $B_{439}$ and 16\,s in
$V_{555}$ for the GO-5324 program and 11\,600\,s in $U_{336}$, 1192\,s
in $V_{555}$ and 1676\,s in $I_{814}$ for GO-7379. These images were
reduced and photometered with the {\tt HSTphot}~1.1 package by
\citet{dol00}, following the recommended procedures in the {\tt
  HSTphot}
manual\footnote{\url{http://www.noao.edu/staff/dolphin/hstphot/}}.

\begin{figure}
  \plottwo{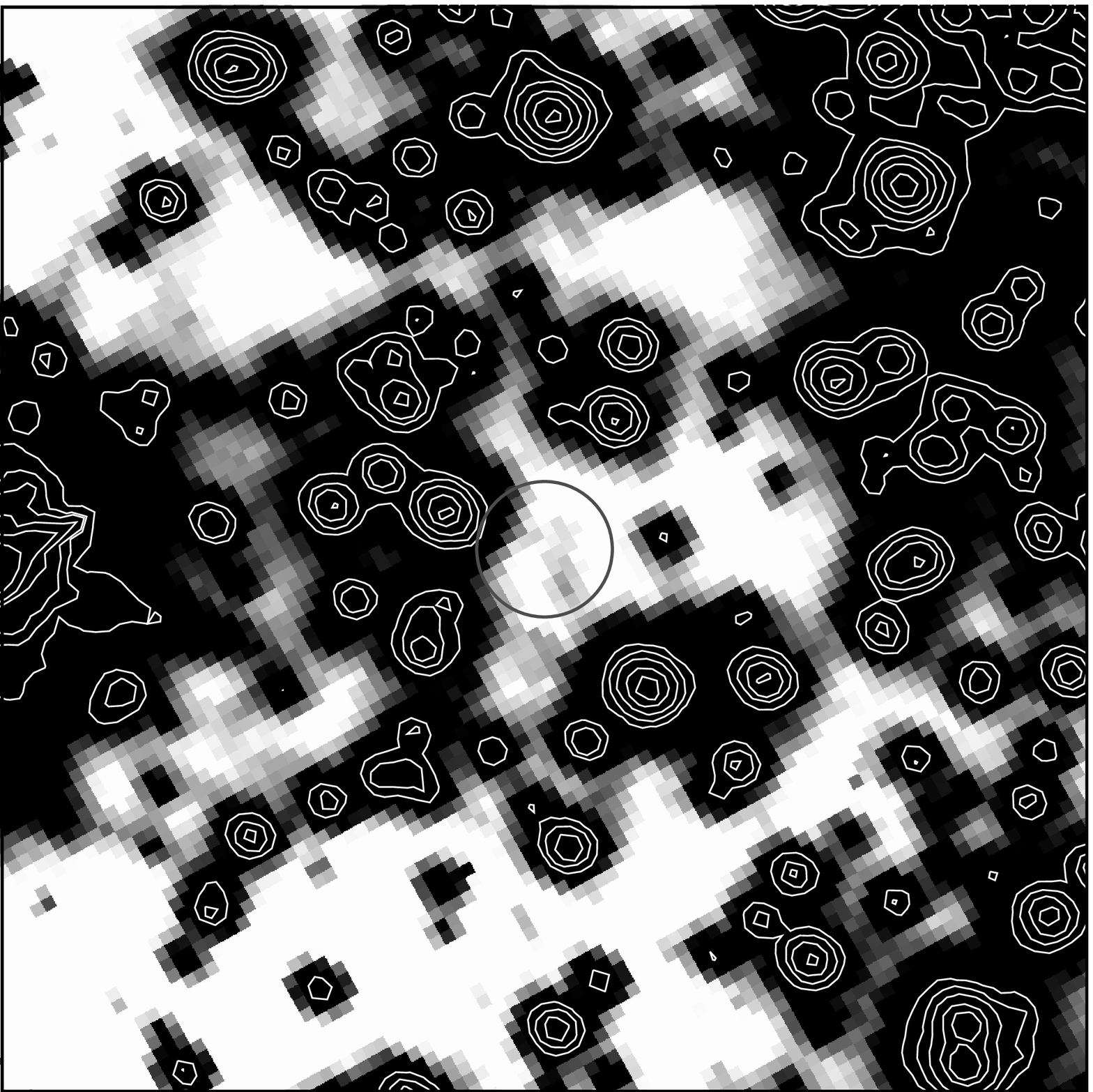}{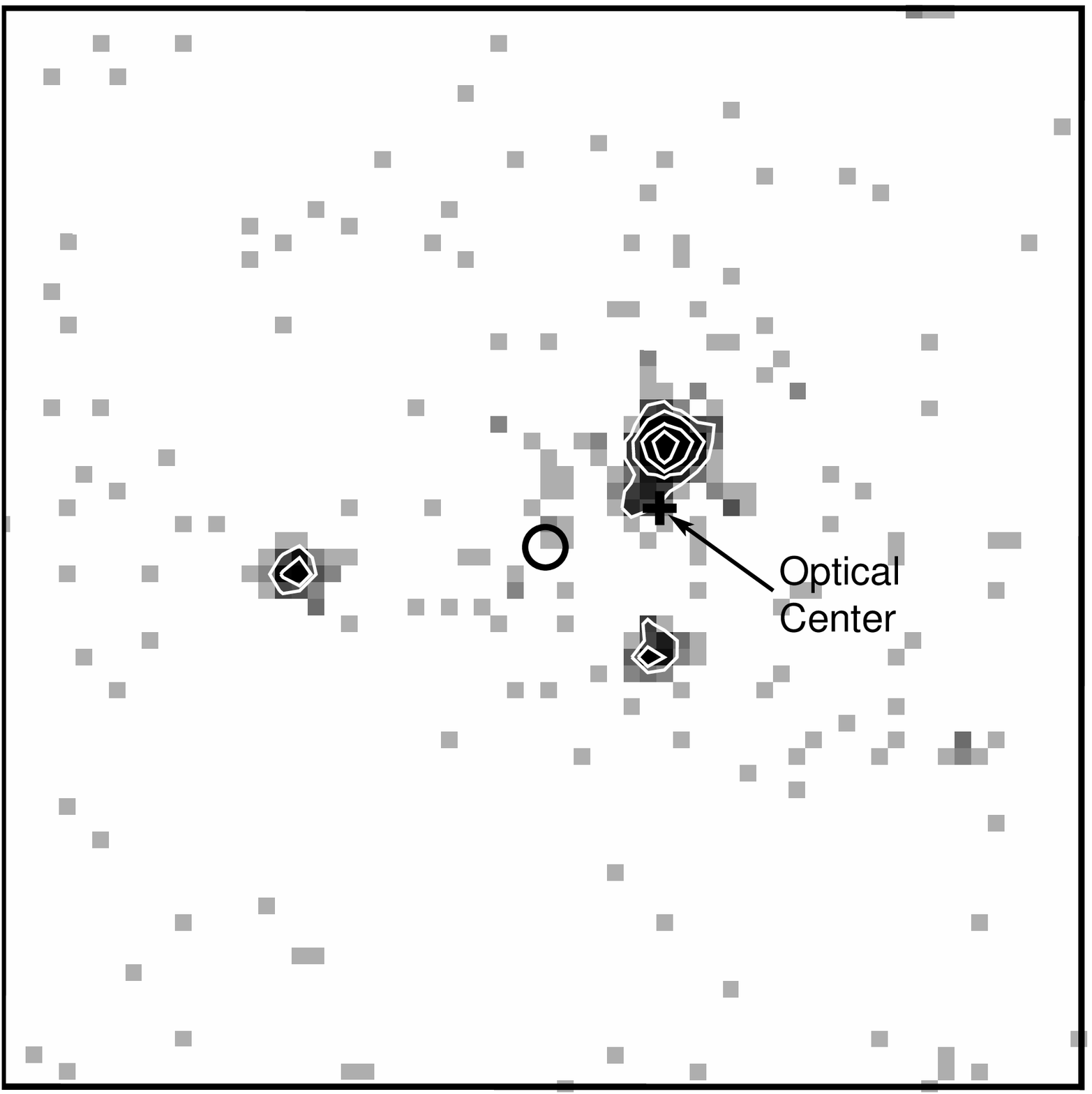} 
  \caption{Optical $V_{555}$-band (left) and X-ray (right) images of
    portions of M30. North is towards the top in each image and east
    to the left. (Left) A $4\asec\times4\asec$ \emph{HST}/WFPC2 F555W
    ($V_{555}$) image with a total exposure of 1044\,s centered on the
    radio position of M30A. The error circle has a radius of
    $0.25\asec$ and is the 95\% confidence region for the position of
    M30A in the optical frame.  A single star is present in the error
    circle, right and below of the center. The countour levels
    correspond to 50, 100, 200, 500, 1000, and 2000 counts.  See
    \S\ref{sec:opt}\ for more details. (Right) A 50\,ks,
    $32\arcsec\times32\arcsec$ {\em Chandra}\ ACIS-S3 0.2$-$3\,keV
    image of the central portion of M30. Light grey pixels indicate a
    single detected photon, while darker pixels indicate two or more
    counts. Contour levels correspond to 5, 15, 50, and 150 counts.
    The error circle has a radius of $0.6\asec$ which is the
    approximate error in the absolute astrometry of the {\em Chandra}
    image.  The astrometrically corrected position of the optical center
    of the cluster as determined by \citet{gwy+98} is marked with a
    black cross. We consider it likely that the $\sim$5 events just to
    the northwest of the nominal pulsar position are from M30A.  See
    \S\ref{sec:xray}\ for more details. }
    \label{fig:images}
\end{figure}

The \emph{HST}/WFPC2 images were placed onto the International
Celestial Reference System (ICRS) for direct comparison against the
position of the pulsar (Table~\ref{tab:M30A}). A detailed description
of this procedure is presented in \citet{bvkh03}.  In short, a
$8\amin\times8\amin$ section of a 4~minute $V$-band image, taken 2000
August 28 with the Wide Field Imager (WFI) at the ESO 2.2~m telescope
at La Silla, was placed onto the ICRS using 58 stars from the USNO CCD
Astrograph Catalog \citep[UCAC;][]{zuz+00}. We fitted for zero-point
position, scale and position angle, and the astrometric solution has
rms residuals of $0.06\asec$ in both right ascension and declination.
This solution was transferred to the WFPC2 images by matching stars on
the WFI with those on the WFPC2 chips, where we corrected the WFPC2
pixel positions for geometric distortion and placed them on a master
frame following the prescription of \citet{ak03}. A total of 99 stars
were used for the GO-7379 dataset (105 for GO-5324), giving rms
residuals of $0.05\asec$ to $0.06\asec$ in both coordinates. The final
($1\sigma$) uncertainty in the astrometric solution of the WFPC2
images is about $0.07\asec$ to $0.08\asec$ in right ascension and
declination. A $4\asec\times4\asec$ section of an image comprising
1044\,s of the $V_{555}$ data from GO-7379 is shown in
Figure~\ref{fig:images}.  The image shows a single star within the
$0.25\asec$ (95\% confidence) error circle centered on the M30A
position.  The star is offset from the pulsar position by $-0.09\asec$
in right ascension and $-0.07\asec$ in declination and is a plausible
companion to M30A.

\notetoeditor{I would like the following two images to appear across
  the full page, so that each image takes up approx 1 column width if
  possible.}

{\tt HSTphot} identified and measured the magnitude of the candidate
companion to M30A in 13 of the 48 individual $V_{555}$ images from
GO-7379, but failed to detect it in the $U_{336}$ and $I_{814}$
images. The S/N ratio of the $V_{555}$ detections is very low, roughly
between 2$-$5, and hence the uncertainties in the magnitude
determinations are large (typically 0.4 magnitudes). The average
$V_{555}$ magnitude of the candidate from the 13 detections is
$23.8\pm0.1$. We folded the magnitude measurements at the orbital
period of the pulsar using the ephemeris in Table~\ref{tab:M30A} but
found no significant variability, possibly due to the large
uncertainties in the magnitudes.  We note that the non-detections
occur roughly uniformly in orbital phase and therefore cannot be
explained exclusively by orbital phase dependent variability.  These
measurements are consistent with visual inspection of the candidate
position on the individual $V_{555}$ images.  The non-detections of
the candidate companion in $U_{336}$ and $I_{814}$ place the object on
or bluewards of the cluster main sequence.

Whether or not this candidate is the actual companion to M30A, it is
unlikely that the true companion is similar to that of
PSR~J1740$-$5340A in NGC\,6397 \citep{fpd+01}. That star is as
luminous as stars at the main sequence turn off, while the $U_{336}$
magnitude of the candidiate companion to M30A is at least 5 magnitudes
fainter than that. If the companion is a main sequence star, it is
more likely comparable to PSR~J0024$-$7204W in 47~Tucanae
\citep{clf+00}, whose binary companion was recently identified in
\emph{HST} observations on the basis of large-amplitude variability
which matched the orbital period and phase of the pulsar
\citep{egc+02}.  Since M30A has similar orbital characteristics to
47~Tuc~W, it is plausible that the companion to M30A shows optical
variability as well.  The current $V_{555}$ data, though, rule out
variability at the 1.6\,mag level as seen from 47~Tuc~W.  A future
series of \emph{HST}/ACS images might unambiguously identify our
candidate as the companion to M30A on the basis of lower amplitude
variability or detect a more likely counterpart.

We have followed the recommendation of \citet{gwy+98} and matched
their astrometry against our astrometric solution of the GO-5324
dataset. We find that their reference star (No.~3611) has an absolute
position of $\alpha_{2000}=21^{\rm h}\;40^{\rm m}\;22\fs314$ and
$\delta_{2000}=-23\degrees\;10\amin\;40\farcs10$. We have furthermore
matched the \citet{gwy+98} positions of the 40 stars in their Table~1
against our positions and solved for zero-point offset, scale and
rotation. The \citet{gwy+98} frame had to be shifted by $-1.25\asec$
and $2.27\asec$ in R.A. and in Decl., scaled by a factor of 0.988 in
both axes and rotated by $-0.76\degrees$ around the new position of
the reference star, to result in our astrometric frame. The rms
residuals of the transformation were about $0.006\asec$ in both
coordinates. In our absolute astrometric frame the \citet{gwy+98}
position of the M30 centroid has $\alpha_{2000}=21^{\rm h}\;40^{\rm
  m}\;22\fs16$ and $\delta_{2000}=-23\degrees\;10\amin\;47\farcs6$. We
use this position for the M30 cluster center throughout the paper.

\subsection{X-ray Observations\label{sec:xray}}

We have also attempted to identify pulsar M30A in a 50\,ks {\em
  Chandra}\ ACIS-S3 observation centered on the core of M30 and taken
2001 November 19 (OBSID~2679).  We used the
CIAO\footnote{\url{http://asc.harvard.edu/ciao}} software package
(v3.01 with CALDB v2.23) to apply the most up-to-date aspect, charge
transfer inefficiency, and gain map corrections to produce a new
Level=2 events file from the archival Level=1 events.  From that file
we kept all events with energies in the range 0.3$-$3\,keV in order
to produce the 32\arcsec$\times$32\arcsec\ image of the cluster core
shown in Figure~\ref{fig:images}.  Just to the north-west of the
nominal pulsar position there appears to be a weak ($\sim$5 counts
with energy $<$3\,keV) source.  While this ``source'' was not detected
by the WAVDETECT algorithm, given that several eclipsing MSPs have
been identified in X-rays \citep[e.g.][]{gch+02,sgk+03}, and that the
timing position of M30A is within the $\sim$0\farcs6 absolute
astrometric error circle for {\em Chandra}, it is plausible that the
observed X-ray flux originates from M30A.

If we assume a blackbody model (using
PIMMS\footnote{\url{http://heasarc.gsfc.nasa.gov/Tools/w3pimms.html}})
with a single $kT\sim0.22$\,keV \citep[as in][]{gch+02} and a $N_H$
for M30 of $3.5\times10^{20}$\,cm$^{-2}$ \citep{dl90}, the five
measured events in the 0.3$-$3\,keV band correspond to an equivalent
unabsorbed flux from the source of
$\sim5\times10^{-16}$\,ergs\,cm$^{-2}$\,s$^{-1}$, which at the
distance of M30 corresponds to a luminosity of
$\sim5\times10^{30}$\,ergs\,s$^{-1}$.  This is comparable to the X-ray
luminosities found for the MSPs in 47~Tucanae and NGC~6397 which are
believed to be due to emission from hot polar caps on the NSs
\citep{gch+02}.  Alternatively, the emission could be due to thermal
bremsstrahlung (TB) in the hot ionized wind of the companion star
where the radio eclipses take place.  While a TB\ model with
$T\sim1$\,keV yields an almost identical luminosity of
$\sim6\times10^{30}$\,ergs\,s$^{-1}$, the implied emission measure
would be ${\rm EM} \sim 2\times10^{54}$\,cm$^{-3}$.  If we assume the
plasma to be fully ionized and uniformly distributed in a spherical
region of diameter $R_E$, the total volume of the plasma is $\sim
1.3\times10^{32}$\,cm$^3$ and the corresponding electron density is
$\sim 1\times10^{11}$\,cm$^{-3}$.  This value is $\sim3000$ times
larger than the measured electron density near the eclipse boundaries
(\S\ref{sec:eclipse}).  Since the eclipsing material is being
constantly replenished (based on the fact that size of the eclipsing
region is much larger than even the Roche lobe radius of the
companion) and since we measure no dispersive delays outside of the
orbital phases corresponding to the eclipse region, the very large
electron densities implied by TB\ effectively rule it out as the
source of the X-ray emission.  An additional argument against the TB\ 
model comes from the fact that the isolated MSPs in 47~Tucanae show
X-ray emission yet have no companion stars to generate the required
hot ionized wind.

%
%
%
%

\section{Pulsar J2140$-$23B (M30B)}
\label{sec:m30b}

For each of the timing observations, we searched the data using the
same techniques described in \S\ref{sec:search} both for new pulsars and
for a re-appearance of M30B.  In addition, we performed limited
folding searches of 20-min to 1-hr data segments for M30B over a range
of periods centered on its discovery spin period and a range of
plausible period derivatives that would be caused by the accelerations
of the possible orbits discussed in \S\ref{sec:m30borbits}.  In no
case have we either definitively re-detected M30B or confirmed any
additional pulsars in the cluster.

Our lack of subsequent detections of pulsar M30B can be explained in
one of two ways.  The most likely explanation is simply that M30B has
a significantly lower average flux density than M30A, and that its
discovery was the result of an extreme diffractive scintillation
event.  Given the exponential distribution of diffractive
scintillation events, however, and the implied low probability of
observing extreme scintillation events, it is unlikely that M30B has a
flux density more than a factor of $\sim$2 below that of M30A.  This
implies that if diffractive scintillation is the cause for our
non-detections in blind searches, then the pulsar would be visible in
much of our data if we could search for it optimally (i.e.~folding on
the correct orbital ephemeris).  We note that M30A is not detectable
via blind searches in approximately half of our observations.

The other possibility for the non-detections of M30B is unusual
eclipses.  It could be that M30B has an intrinsic luminosity similar
to or even greater than that of M30A, but its companion causes
irregular and possibly long-term ``eclipses'' of the pulsed signal in
the same manner as the GC MSPs PSR~B1744$-$24A in Terzan~5
\citep{ljm+90b,nt92} and PSR~J1740$-$5340 in NGC~6397 \citep{dpm+01a}.

\subsection{Orbit Constraints\label{sec:m30borbits}}

\begin{figure}
  \plotone{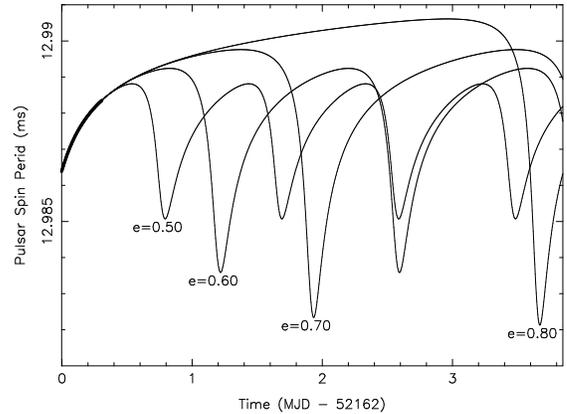}
  \caption{
    The predicted time dependence of the measured spin period of M30B
    given four representative Keplerian orbital solutions from the
    family shown in Fig.~\ref{fig:M30Bparams}.  The thick black line
    shows the measured spin period during the discovery observations
    of 2001 September 9.
    \label{fig:M30Borbits}}
\end{figure}

Without post-discovery detections of M30B we are currently unable to
unambiguously determine the orbital parameters of the system.
However, examination of the spin-frequency behavior during the more
than seven hours that the pulsar was visible in the 2001 September 9
discovery observation immediately led us to conclude that we could not
fit a sinusoid to the frequency behavior (see
Fig.~\ref{fig:M30Borbits}), and hence that the orbit was significantly
eccentric.

We measured 26~TOAs from the BCPM1 and BCPM2 discovery observations
and used {\sc TEMPO} to fit a timing model consisting of the DM of
M30B and a Taylor expansion of the measured pulsation frequency in
time.  Reasonable solutions providing ``white'' timing residuals
required fits to the initial spin frequency ($f$) and at least the
first four frequency derivatives ($\dot f$, $\ddot f$, $f^{(3)}$,
and $f^{(4)}$).  The best 4-derivative solution, at an MJD(UTC)
epoch of 52161.997395, and a DM of 25.09$\pm$0.12\,pc\,cm$^{-3}$, is
\begin{eqnarray*}
  \label{tab:M30B}
  f       &=& 77.003746(14)            \,{\rm Hz},         \\
  \dot f  &=& -8.557(59)\times10^{-7}  \,{\rm Hz\,s^{-1}}, \\
  \ddot f &=& 5.68(16)\times 10^{-11}  \,{\rm Hz\,s^{-2}}, \\
  f^{(3)} &=& -4.06(25)\times 10^{-15} \,{\rm Hz\,s^{-3}}, \\
  f^{(4)} &=& 1.79(19)\times 10^{-19}  \,{\rm Hz\,s^{-4}},
\end{eqnarray*}
and provided RMS residuals of $\sim$53\,$\mu$s and the pulse profile
shown in Fig.~\ref{fig:profiles}.

\begin{figure}
  \plotone{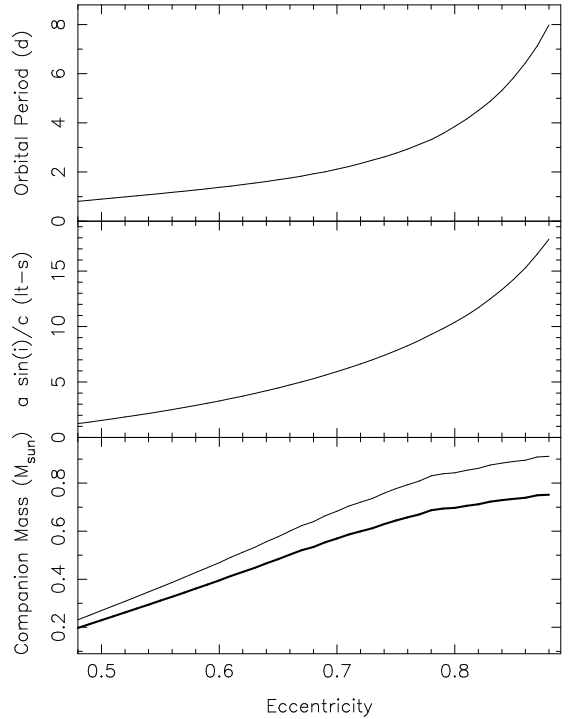}
  \caption{
    The family of possible Keplerian orbital solutions for M30B based
    on {\sc TEMPO} fits to 26~TOAs determined during the discovery
    observation.  The 95\% confidence lower-limit to the eccentricity
    is $e\ga0.52$ (see \S\ref{sec:m30borbits}).  For $P_{orb}$ (top)
    and $a\sin(i)/c$ (middle), typical uncertainties are 10$-$20\% in
    the plotted value.  In the bottom panel, the lower line shows the
    minimum companion mass ($i=90\degrees$) while the upper line
    represents the mass corresponding to the median inclination,
    $i=60\degrees$, assuming random inclinations.  In all cases we
    have assumed a pulsar mass of $m_1=1.4\,\msun$.
    \label{fig:M30Bparams}}
\end{figure}

Given the timing solution above, we applied the orbit-inversion
technique described by \citet{jr97} to convert the polynomial-based
solution into a family of Keplerian orbital solutions as a function of
orbital eccentricity, $e$\ (assuming that the intrinsic $\dot f$ for
the pulsar is negligible during the observation).  A lack of physical
solutions from the technique provided a rough lower limit to M30B's
eccentricity of $e\gtrsim 0.45$.  Using the inversion-based family of
solutions as starting points at a series of trial eccentricities, and
setting the DM to the value measured earlier, we used {\sc TEMPO} to
compute $\chi_\nu^2$ surfaces (with $\nu$=22 degrees-of-freedom) as a
function of $P_{orb}$ and $x$, by allowing the software to fit for
$f$, the angle of periastron ($\omega$) and the time since periastron
($T_0$).  The values of $P_{orb}$ and $x$ at the resulting
$\chi_\nu^2$ minima are plotted in Fig.~\ref{fig:M30Bparams}, and have
typical errors in each parameter of 10$-$20\%.  This process provided
a 95\% confidence lower limit to the eccentricity of $e\ge0.52$.
Orbits with $e\ge0.55$ provided essentially perfect fits to the data
with $\chi_{\nu}^2/\nu\sim1.0$ and white-noise-like residuals with an
RMS of $\sim$30\,$\mu$s.  The spin-period behavior of four
representative orbital solutions is plotted in
Fig.~\ref{fig:M30Borbits} as well as the measured spin-period behavior
during the discovery observation.

The family of orbital solutions that we have determined implies that
M30B is part of a relativistic orbital system with a companion most
likely in the mass range 0.25$-$1\,\msun.  If the companion is a white
dwarf (WD), the advance of periastron ($\dot\omega$) due to general
relativity will be several tenths of a degree per year and would be
measurable (as will, therefore, the total system mass, $M=m_1+m_2$) to
a high degree of precision within a year of monthly timing
observations of a quality similar to the discovery observation.  It is
also possible, if such timing measurements can be made on a consistent
basis, that additional post-Keplerian orbital parameters, such as the
Einstein $\gamma$, will be measurable in M30B within several years.
Such measurements would provide one of the few accurate mass
determinations of a MSP which is important for testing models of NS
spin-up via accretion.  Most models predict the accumulation of
$\gtrsim$0.1\,\msun\ of material onto the NS during the spin-up
process \citep[][and references therein]{pk94}, and recent
observations have begun providing evidence for the case
\citep*[e.g.][]{nss02}.

In the current pulsar catalog, there are only four other GC binary
pulsars with $e>0.1$.  The NS$-$NS system in M15, PSR~B2127$-$11C
\citep[$P=30.5$\,ms, $P_{orb}=0.335$\,d, $e=0.681$;][]{pakw91} was
almost certainly created in an exchange interaction near the core of
the cluster that replaced a lower-mass MS or WD companion of a NS with
another NS \citep{ps91, sp93b}.  This interaction also placed the
system as a whole in a highly eccentric orbit around the
center-of-mass of M15, which explains its location near the outskirts
of the cluster today.  An exchange interaction also provides the most
likely explanation for the eccentricity of PSR~B1802$-$07 in NGC~6539
\citep[$P=23.1$\,ms, $P_{orb}=2.62$\,d, $e=0.212$;][]{dbl+93} --- a
system possibly very similar to M30B --- and a possible explanation
for the eccentricity of PSR~B1516$+$02B in M5 \citep[$P=7.95$\,ms,
$P_{orb}=6.86$\,d, $e=0.138$;][]{awkp97} as well.  \citet{rh95} have
argued, however, that M5B's eccentricity may be the product of many
distant interactions with other stars near the clusters core over
several Gyr.  The fourth cluster binary pulsar with a large
eccentricity is the recently discovered PSR~J1750$-$37 in NGC~6441
\citep[$P=111.6$\,ms, $P_{orb}=17.3$\,d, $e=0.71$;][]{pdm+01}, which
may be a very similar system to M30B, albeit with a much slower spin
period.  Hypotheses of its origin await determination of its position
with respect to the cluster center, but with such a large
eccentricity, an exchange encounter seems likely.

We can estimate the time required for a pulsar with an initially
near-circular orbit in the core of M30 to accumulate an eccentricity
of $e\ge0.5$ by using Equation~5 from \citet{rh95}.  With a
one-dimensional velocity dispersion for the cluster core of
$v_\ell(0)$=9.4$\pm$2.5\,km\,s$^{-1}$ \citep{gpw+95}, and the number
density of stars in the core of $\sim1.6\times10^5$\,pc$^{-3}$
\citep{gwy+98}, we estimate $t_{e\ge0.5} \simeq 7
P_{orb,d}^{-2/3}$\,Gyr, where $P_{orb,d}$ is the orbital period of
M30B in days.  Therefore, if M30B resides in the core of M30, it is
possible that its eccentricity is due to many distant interactions.
It is interesting to note that for M30A, which probably does reside in
the core of M30, the time required for multiple distant interactions
to produce an eccentricity that would be detectable in our current
measurements ($e\sim2\times10^{-4}$) is $\sim$2$-$3\,Gyr.  This means
that M30A is either younger than $\sim$2$-$3\,Gyr, or the mechanism
described by \citet{rh95} is not as efficient in M30 as we calculate.

Should M30B reside outside of the cluster core, that may indicate that
its high eccentricity was caused by an exchange interaction similar to
that which could have produced PSR~B1802$-$07 and almost certainly did
produce the recently discovered eclipsing MSP in the outskirts of
NGC~6397, PSR~J1740$-$5340 \citep{dpm+01a}.  If the exchange
interaction replaced a low-mass WD with a MS star, optical
observations should allow its identification, once a timing position
has been measured.  Such a MS or even post-MS companion could help to
explain our lack of detections of the pulsar as well.  If the
companion star experiences significant though possibly erratic mass
loss, the pulsed flux from M30B could be eclipsed or even quenched
entirely for periods at a time, just as observed from
PSR~J1740$-$5340.  If this is the case, higher frequency observations
of the cluster may allow us to peer through the wind of the companion
and see the pulsar consistently.

If the companion to M30B is a MS or post-MS star, though, orbital
circularization due to a hydrodynamical mechanism \citep{tas95} may
occur on timescales much shorter than the $\sim$10$^{8-10}$\,yr age of
the system we would expect if it was created via an exchange
interaction.  \citet{tas95} calibrated a theoretical relation for
hydrodynamical orbit circularization by comparing the shortest orbital
periods of MS binaries with eccentric orbits in several stellar
clusters of different ages.  He determined the time for
circularization in years to be roughly $t_{circ} \sim
1.5\times10^{(7-N/4)}P_{orb,d}^{49/12}$, where the fitting parameter
$N\sim8.3$ for stellar systems of age $>$1\,Gyr and $P_{orb,d}$ is the
orbital period in days.  For the family of orbital solutions we have
found for M30B, this corresponds to $t_{circ} \sim 10^{5-9}$\,yr, and
implies either that the companion star is a WD, or that the
interaction that exchanged a MS companion into the system happened
within the last 10$^9$\,yrs, or significantly more recently if the
orbital period of M30B is $\la$6\,d.  We note, though, that the
\citet{tas95} calculations assume a binary system of two MS stars and
almost certainly need modification for NS$-$MS systems.

\section{Conclusions}
\label{sec:conc}

We have used the Green Bank Telescope to discover two binary MSPs in
the core-collapsed GC M30.  One of these systems is a member of the
rapidly growing class of eclipsing MSPs and the other seems to be in a
highly eccentric and relativistic orbit.  Higher sensitivity
observations of M30 in the near future --- for instance, using all of
the available bandwidth provided by the GBT's 20-cm receiver ---
should allow us to monitor M30A for long-term variations in its
orbital parameters as has been seen in other eclipsing MSPs
\citep*[e.g.][]{aft94} and enable us to consistently detect and time
M30B.  Additional observations at other radio frequencies will allow
us to probe the eclipse region of M30A to a greater degree, and will
hopefully allow us to constrain the mechanism behind the eclipses
themselves.

It seems clear that as in the case of most other GCs
\citep[e.g.][]{clf+00}, we are only seeing the most luminous pulsars
contained in M30.  Given the extreme scintillation events we have
witnessed from M30A and B, we consider it quite likely that the
improved observations mentioned above will uncover several additional
pulsars in M30 in the years ahead.

{\em Acknowledgements} We would like to thank the anonymous referee
for comments which significantly improved the structure of the paper,
Mallory Roberts and Maxim Lyutikov for useful discussions, and Frank
Ghigo, Glen Langston, Toney Minter, and Richard Prairie for assistance
with the observations.  SMR\ acknowledges the support of a Tomlinson
Fellowship awarded by McGill University.  IHS\ holds an NSERC\ 
University Faculty Award and is supported by a Discovery Grant.  CGB
is supported by the Netherlands Organization for Scientific Research.
The computing facility used for this research was funded via a New
Opportunities Research Grant from the Canada Foundation for
Innovation.  Additional support is from an NSERC\ Discovery Grant and
from NATEQ.  The National Radio Astronomy Observatory is a facility of
the National Science Foundation operated under cooperative agreement
by Associated Universities, Inc.  This research has made extensive use
of NASA's Astrophysics Data System (ADS) and High Energy Astrophysics
Science Archive Research Center (HEASARC).  Data from European
Southern Observatory telescopes was obtained from the ESO/ST-ECF
Science Archive Facilities.




\end{document}